\documentclass[12pt]{article}
\usepackage{fullpage}
\usepackage{amsmath}
\usepackage{amssymb}
\usepackage{amsthm}
\usepackage{graphicx}
\usepackage{fancyhdr}
\usepackage{feynmf}
\usepackage{cite}
\usepackage{hyperref}

\usepackage{color}
\definecolor{darkblue}{rgb}{0,0,0.5}

\begin{document}

\title{{\bf Another Unorthodox Introduction to QCD and now Machine Learning}}
\author{Andrew J.~Larkoski\footnote{larkoski@reed.edu}\\ \\\small{Physics Department}\\\small{Reed College}}
\date{}
\maketitle

\begin{abstract}
\noindent These are lecture notes presented at the online 2020 Hadron Collider Physics Summer School hosted by Fermilab.  These are an extension of lectures presented at the 2017 and 2018 CTEQ summer schools in \cite{Larkoski:2017fip} and still introduces perturbative QCD and its application to jet substructure from a bottom-up perspective based on the approximation of QCD as a weakly-coupled, conformal field theory.  With machine learning becoming an increasingly important tool of particle physics, I discuss its utility exclusively from the biased view for increasing human knowledge.  A simple argument that the likelihood for quark versus gluon discrimination is infrared and collinear safe is presented as an example of this approach.  End-of-lecture exercises are also provided.
\end{abstract}

\clearpage

\section{Lecture 1: QCD as a Weakly-Coupled Conformal Field Theory}

Hello!  I'm Andrew Larkoski, a professor at Reed College in Portland, Oregon, USA, and I'm excited to lecture at this year's Hadron Collider Physics Summer School.  I have been tasked with discussing hadronic jets, which are collimated streams of particles created from the dynamics of quantum chromodynamics (QCD) at high energies.  In this first lecture, I'll introduce jets from a ``bottom-up'' perspective, forgoing any explicit discussion of the fundamental QCD Lagrangian from which jets arise as an emergent phenomena.  This will be similar in approach to the recommended reading of my lectures from the CTEQ summer schools from a few years ago, Ref.~\cite{Larkoski:2017fip}.  I've also been asked to discuss machine learning for jet physics.  Now, as a theoretical physicist with research interests in calculations, I won't be discussing neural network architecture, programming, computer science, etc., but will instead propose the following.  Regradless of your individual thoughts of machine learning, the manner in which data is input and output of a machine suggests a novel way of thinking about fundamental problems in particle physics generally, and jet physics specifically.  I'll discuss this way of thinking and how it can be used to derive and learn about very general results in the next lecture.

For this lecture, let's understand what QCD is in the first place.  My introduction will be very different (likely) than you've seen before.  To define QCD, jets and their consequences, we will only make two assumptions or axioms from which everything in these lectures follows.  They are:
\begin{enumerate}

\item At high energies, QCD is an approximately scale-invariant quantum field theory.  This means that, to first approximation, the coupling of QCD, $\alpha_s$, is constant, independent of energy.

\item Not only is $\alpha_s$ (approximately) constant, but further $\alpha_s$ is small; formally we assume that $\alpha_s \ll 1$.  This means that the degrees of freedom in the Lagrangian of QCD, quarks and gluons, are good quasi-particles for actually describing the physics of QCD.

\end{enumerate}

These assumptions are sufficient to write down matrix elements for some simple processes.  Actually, we'll just write down the probability density functions, because ``matrix element'' (implicitly) assumes specific interactions and particle properties, but our axioms say nothing explicit about interactions.  By axiom 2, we can ask questions about the dynamics of quarks and gluons.  So, we'll ask: what is the probability for a quark to emit a gluon?  With a Lagrangian and Feynman rules, we would want to calculate the (squared) diagram:
\begin{equation}
P_{qg\leftarrow q} = \left|
\raisebox{-1.2cm}{\includegraphics[width=5cm]{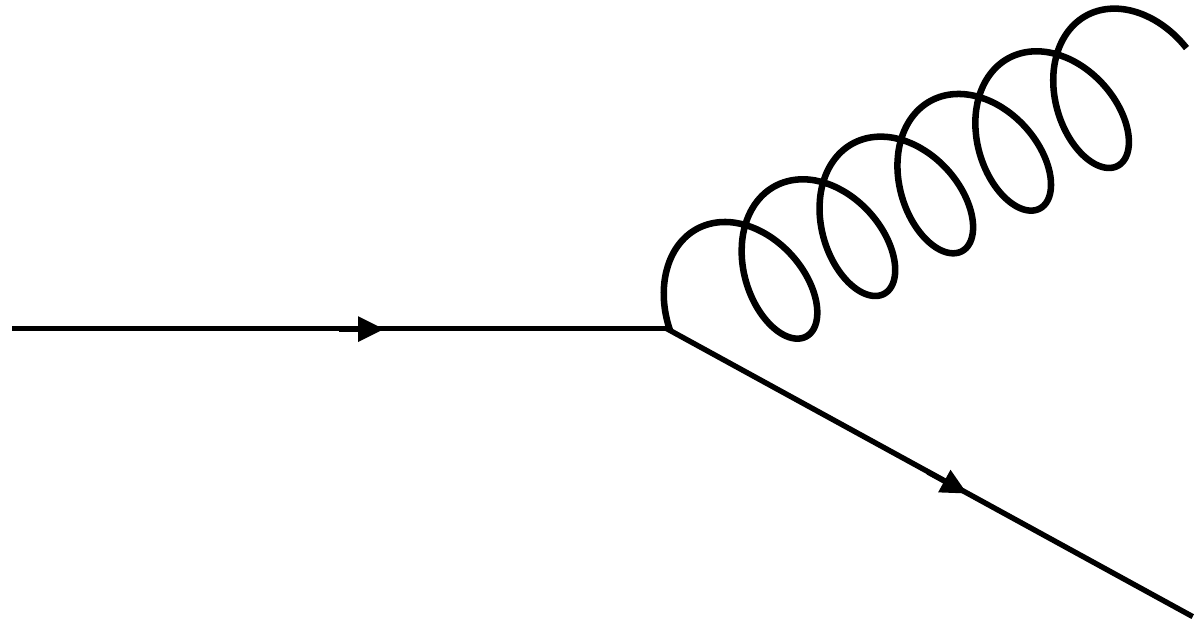}}
\right|^2\,.
\end{equation}
But we don't have Feynman rules, so we have to use our axioms.

We have a few things to establish before our axioms, however.  The probability for gluon emission will, in general, depend on the four-momentum of the gluon.  So, we need to identify the space on which the probability $P_{qg\leftarrow q}$ is defined.  Then, given that space, we can ask what constraints scale invariance imposes.  That is, we need to identify the degrees of freedom of the emitted gluon.

Let's start with the gluon's four-momentum written as:
\begin{equation}
p_g = (E,p_x,p_y,p_z)\,.
\end{equation}
The gluon is massless, so demanding that this momentum be on-shell requires $E = |\vec p|$, or that the momentum can be expressed in sphereical coordinates as
\begin{equation}
p_g = E(1,\sin\theta \cos\phi,\sin\theta\sin\phi,\cos\theta)\,.
\end{equation}
Here, $\theta$ is the quark-gluon opening angle and $\phi$ is the azimuthal angle about the quark:

\begin{center}
\includegraphics[width=5cm]{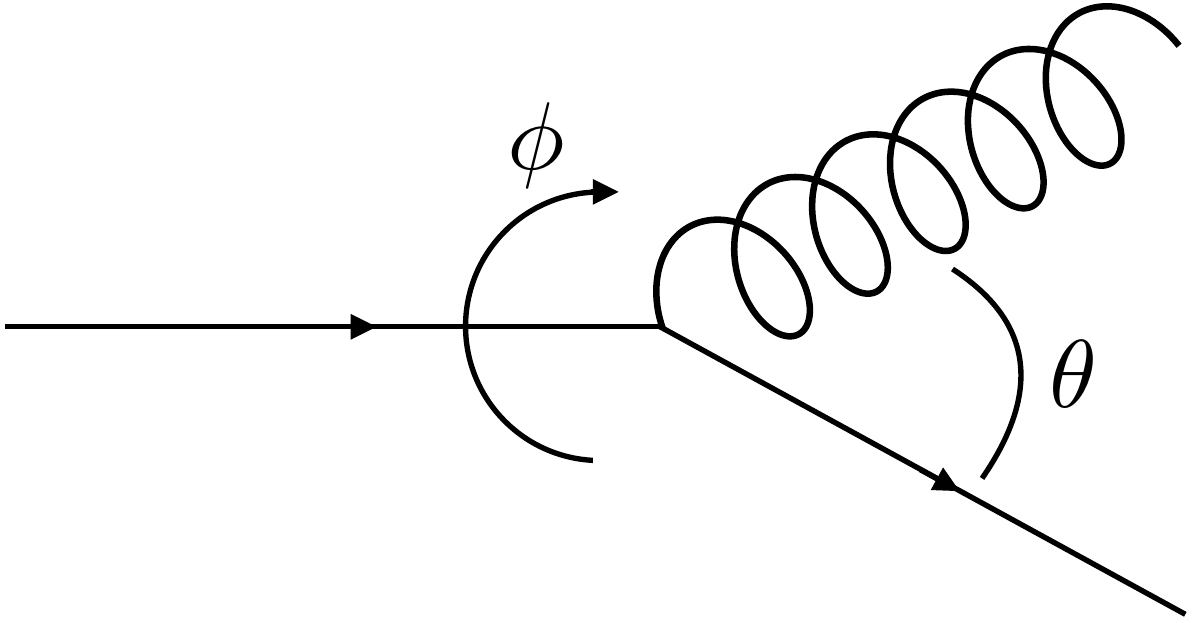}
\end{center}
Additionally, the total quark plus gluon energy is fixed.  Calling this total energy $E_\text{tot}$, we can write $E = zE_\text{tot}$, for some fraction  $z\in[0,1]$.  Additionally, from the formulation of our assumptions and the problem at hand, there was nothing special about the azimuthal angle $\phi$: the physics is independent of $\phi$.  Thus, we can fix $\phi$ to a convenient value, say, $\phi = 0$.  (That is, we assume that the quark is unpolarized.)  With these identifications, the gluon's momentum is:
\begin{equation}
p_g = zE_\text{tot}(1,\sin\theta,0,\cos\theta)\,.
\end{equation}
We had assumed that the final state quark's momentum is along the $z$-axis, so it is
\begin{equation}
p_q = (1-z)E_\text{tot}(1,0,0,1)\,.
\end{equation}
We have therefore identified the gluon's relevant degrees of freedom to be the energy fraction $z$ and angle $\theta$.  The splitting probability is:
\begin{equation}
P_{qg\leftarrow q} = p(z,\theta)\, dz\, d\theta\,,
\end{equation}
for some other function $p(z,\theta)$.

There are a few things we can immediately say about the function $p(z,\theta)$.  $\alpha_s$ is the QCD coupling; as such, it controls the strength with which quarks and gluons interact with one another.  Thus, this function, as it describes the probability of emission of a gluon off of a quark, is proportional to $\alpha_s$:
\begin{equation}
p(z,\theta) \propto \alpha_s\,.
\end{equation}
Actually knowing QCD we can say more, so we might as well add it.  The actual factors that come with $\alpha_s$ are:
\begin{equation}
p(z,\theta)\propto \frac{2\alpha_s}{\pi}C_F\,,
\end{equation}
where $C_F$ is the fundamental quadratic Casimir of the SU(3) color symmetry of QCD.  Again, our assumptions can't tell us about these factors, but they won't qualitatively change the picture we are developing (yet).  In QCD, $C_F = 4/3$, and so $2\,, \pi\,, C_F \sim {\cal O}(1)$, so truly $\alpha_s$ is what is controlling the coupling of quarks and gluons, by the assumption that $\alpha_s \ll 1$.  By the way, $C_F$ is a measure of how quarks and gluons share the three colors of QCD.

Next, we'd like to determine the dependence of $p(z,\theta)$ on the energy fraction $z$ and angle $\theta$.  To do this, we need to think about what our assumption of ``scale invariance of QCD'' means.  QCD is a quantum field theory, and as such is Lorentz invariant.  Thus, the only quantities that all observers agree on are those that are, well, Lorentz invariant.  If we further state that QCD is scale-invariant, this means that a scaling of all Lorentz-invariant quantities produces the same physics.  That is, probability distributions of Lorentz invariant quantities should be further invariant to scale transformations.

Given our quark-gluon system, the only Lorentz invariant we can construct is the dot product of their momenta:
\begin{equation}
p_q\cdot p_g = z(1-z)E^2_\text{tot}(1-\cos\theta)\,.
\end{equation}
This is Lorentz invariant by construction and scale invariance means that the scaling $p_q\cdot p_g \to \lambda\, p_q\cdot p_g$, for any $\lambda>0$, leads to identical physical phenomena.  In general, this scaling is not simply implemented on the energy fraction or angle, but there is a limit in which it is simple.  If the gluon has low energy or is soft so that $z\ll 1$ and is nearly collinear with the quark so that $\theta \ll 1$, note that
\begin{align}
&z(1-z)\xrightarrow{z\ll 1} z\,, &1-\cos\theta \xrightarrow{\theta\ll 1} \frac{\theta^2}{2}\,.
\end{align}
Then, in this double limit, the dot product is
\begin{equation}
p_q\cdot p_g \xrightarrow{z\ll 1\,, \theta\ll 1} z\theta^2\frac{E_\text{tot}^2}{2}\,.
\end{equation}
Thus, the soft and collinear limit corresponds to this dot product expressed as a power law function in both $z$ and $\theta$.  A scaling of $p_q\cdot p_g$ can therefore be accomplished by scaling of either $z$ or $\theta$ (or both).  Then, we identify the scale transformation under which QCD is invariant as
\begin{equation}
z\to \lambda\, z\qquad \text{or} \qquad \theta^2 \to \lambda \, \theta^2\,, \qquad \text{for }\lambda > 0\,.
\end{equation}

Before continuing, note that there is a coordinated scale transformation under which the product $z\theta^2$ is unchanged.  If we scale
\begin{equation}
z\to \lambda\, z\qquad \text{and} \qquad \theta^2 \to \frac{\theta^2}{\lambda}\,, 
\end{equation}
then $z\theta^2\to z\theta^2$.  That is, such a coordinated scale transformation is actually a Lorentz transformation, a boost long the direction of the quark's momentum.  Lorentz invariance states that if energies increase, angles must decrease to ensure that dot products are unchanged.

Now, if scale transformations are implemented by independent scalings $z\to\lambda_1\, z$ and $\theta^2 \to \lambda_2\, \theta^2$, for $\lambda_1,\lambda_2>0$ and the probability $P_{qg\leftarrow q}$ must be unchanged under this scaling, this uniquely fixes the $z$ and $\theta$ dependence of $p(z,\theta)$ to be:
\begin{equation}
P_{qg\leftarrow q} = p(z,\theta)\, dz\, d\theta = \frac{2\alpha_s}{\pi}C_F\, \frac{dz}{z}\frac{d\theta}{\theta}\,.
\end{equation}
This is invariant to the scalings and therefore satisfies our first assumption about QCD.  Note that for this simple functional form, we had to work in the soft and collinear limit.  While this may seem restrictive, we'll be able to get a lot of mileage out of it.

The first thing to note about this probability distribution is that, in the $z$ or $\theta$ to 0 limits, it diverges.  Actually, it's worse than that: not only does it diverge, but it is not even integrable as $z,\theta\to 0$!  Thus it isn't a ``probability distribution'' at all, because it cannot be normalized.  So how can such a distribution describe some physical process which is non-singular?  Note that the $z$ or $\theta$ to 0 limits are those limits in which the gluon becomes unobservable.  The $z\to 0$ limit is the limit in which the gluon has no energy. A detector, like the calorimeters at ATLAS or CMS, requires a finite energy of the particle to observe it: it never ``sees'' a 0 energy ``hit.''  The $\theta\to 0$ limit is when the gluon is collinear with the quark.  Because it is traveling in the exact same direction as the quark, there is no measurement you can perform to separate them out.  The angular resolution of the cells of the calorimetry at ATLAS and CMS is finite: the particles must be a non-zero angle from one another to be distinguished.

That is, there is no measurement you can perform in the $z\to 0$ or $\theta\to 0$ limit to distinguish a quark emitting one gluon to the case in which the quark emits no gluons.  That is, we call the $z\to 0$ or $\theta\to 0$ limits degenerate, as the physical configuration degenerates to a system with fewer gluons.  But this also points to an extension.  There is no measurement we can perform to distinguish a quark that emitted no gluons:

\begin{center}
\includegraphics[width=3cm]{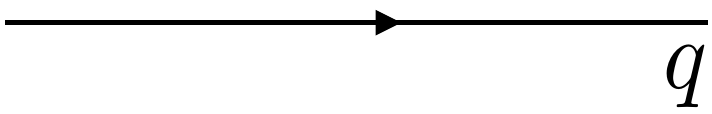}
\end{center}
from a quark that emitted one soft and/or collinear gluon:

\begin{center}
\includegraphics[width=3cm]{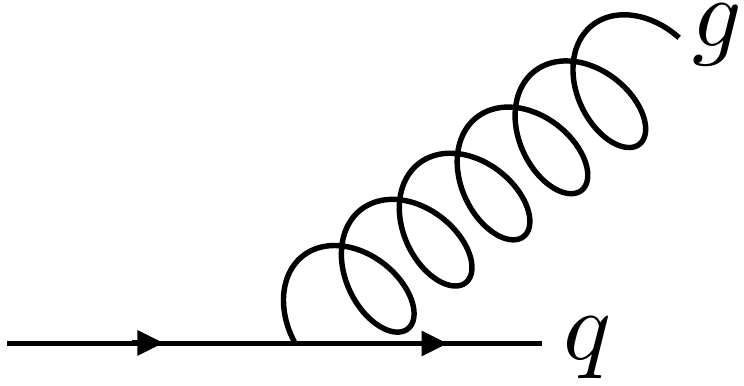}
\end{center}
or two soft and/or collinear gluons:

\begin{center}
\includegraphics[width=3.5cm]{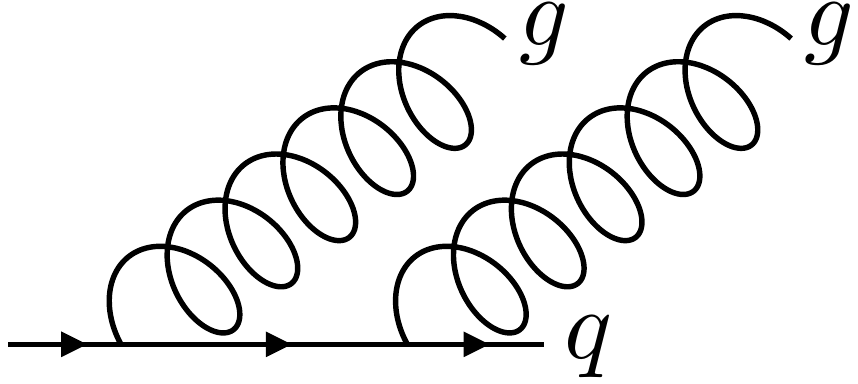}
\end{center}
or any number of soft and/or collinear gluons!  Every one of these configurations of soft and/or collinear gluons plus the quark have a divergent probability and all are degenerate with each other.  So how do we proceed?

Quantum mechanics saves us!  Feynman diagram perturbation theory is a degenerate perturbation theory.  And just like degenerate perturbation theory in quantum mechanics, we only get a finite result by summing over all degenerate configurations.  In quantum field theory, this result is called the KLN theorem \cite{Kinoshita:1962ur,Lee:1964is} (for Kinoshita, Lee and Nauenberg), extending results of Bloch and Nordsieck in quantum electrodynamics from the 1930s \cite{Bloch:1937pw}.  KLN ensures that summing over all individual divergent degenerate configurations produces a finite result.

We'll end this lecture by seeing how this is done.  Let's first go back to the probability distribution
\begin{equation}
P_{qg\leftarrow q}=\frac{2\alpha_s}{\pi}C_F\, \frac{dz}{z}\frac{d\theta}{\theta} = \frac{2\alpha_s}{\pi}C_F\, d\left(\log\frac{1}{z}\right)\,d\left(
\log\frac{1}{\theta}
\right)\,.
\end{equation}
On the right, I just re-expressed the probability as flat logarithmically in $z$ and $\theta$.  Thus, the phase space in the $(\log1/z,\log1/\theta)$ coordinates is a semi-infinite region where $\log1/z,\log1/\theta>0$.  Further, the probability distribution is flat: that is, there is uniform probability for the emission to be anywhere on the plane.  Thinking with degeneracy and KLN in mind, we can generalize this and say that the (arbitrary) gluon emissions off of a quark uniformly fill out the plane.  Representing each emission by a dot, for emissions off of a quark, we might have:

\begin{center}
\includegraphics[width=6cm]{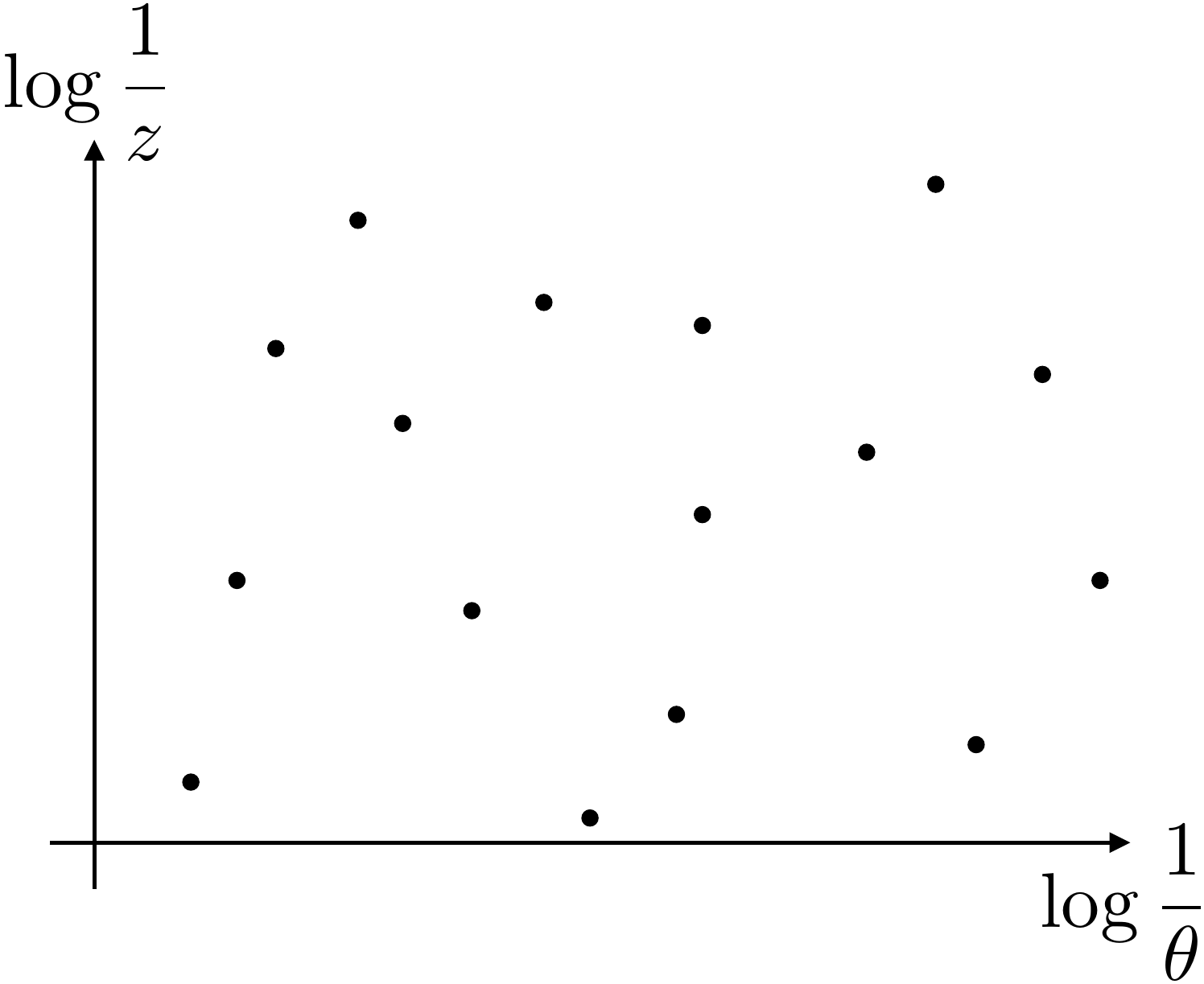}
\end{center}
Uniform in this logarithmic plane means exponentially far apart in ``real'' space, so gluon emissions off of a high energy quark are dominantly soft and/or collinear.  That is precisely what a jet is.

This plane is called the Lund plane, after researchers in Sweden who introduced it for studying jets \cite{Andersson:1988gp}.  Also, this uniform emission distribution is the starting point of modern Monte Carlo event generators and parton showers.  A modern Monte Carlo, like Pythia \cite{Sjostrand:2006za,Sjostrand:2014zea}, Herwig \cite{Bahr:2008pv,Bellm:2015jjp} and Sherpa \cite{Gleisberg:2008ta,Bothmann:2019yzt}, contains significant physics beyond this uniform assumption, such as: running $\alpha_s$ (emissions increase as $z,\theta$ decrease), fixed-order corrections (corrections to $1/z,1/\theta$ distributions), cut off by the scalar hadron masses, etc.  Nevertheless, while simple, this uniform emission phase space has a lot of physics.

Before we do a calculation, I want to orient you in the Lund plane.  Let's draw it again:

\begin{center}
\includegraphics[width=8cm]{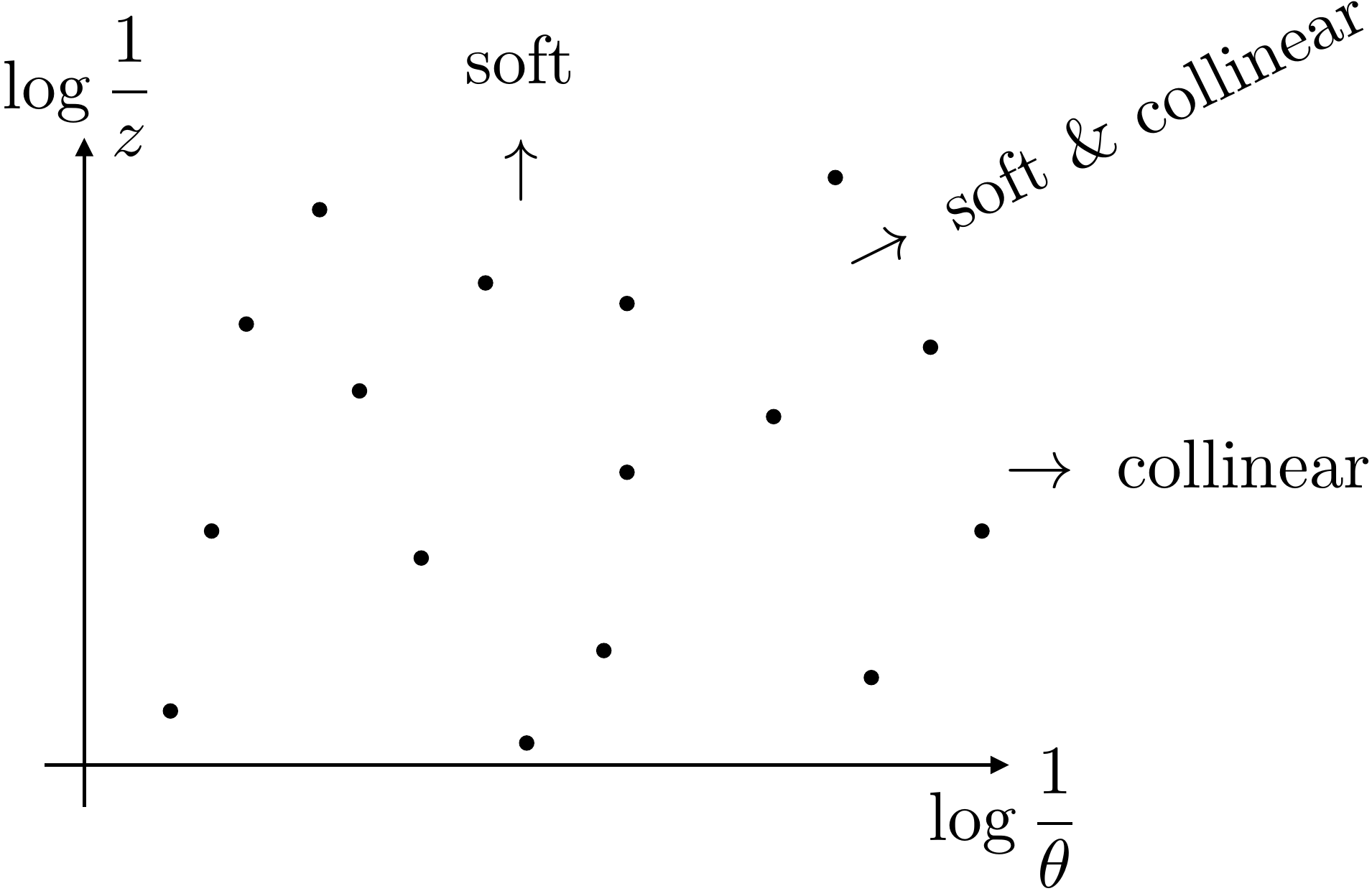}
\end{center}
I've also included regions to guide the eye.  The origin, where $z,\theta\sim 1$, corresponds to high-energy, wide-angle gluon emission.  The degenerate limits live off at $\infty$, with different physical origins for different infinities.  Vertical in the plane, $z\to 0$, is the soft limit, horizontal is the collinear limit $\theta \to 0$, and diagonal is a correlated soft and collinear limit.

With this picture in place, let's now calculate the distribution of a particular observable called an angularity $\tau_\alpha$ \cite{Berger:2003iw,Almeida:2008yp,Ellis:2010rwa,Larkoski:2014uqa}.  The angularity can be defined in the energy fraction $z$/angle coordinates as
\begin{equation}
\tau_\alpha = \sum_{i\in J} z_i\theta_i^\alpha\,,
\end{equation}
where the sum runs over all particles $i$ in the jet $J$ (collections of emissions), and $\alpha > 0$ is an exponent that weights contributions from different angles.  We require that $\alpha > 0$ to ensure that the observable is infrared and collinear safe: as we will see, this means that arbitrarily soft or collinear emissions cannot contribute to the observable (at least not dominantly so).  One important thing to note is that there is no preferred ordering to emissions of a scale-invariant system.  ``Scale invariant'' means that any scale we impose on the system can and will exist in that system.  We will see how measuring the angularities sets one particular ordering.

We had mentioned before that uniform logarithmically means exponentially far in real space.  Because each term in the definition of the angularities is weighted by a product of energy and angle, there will be one emission in the jet that dominates the value of $\tau_\alpha$, and all others will be exponentially small.  So, with one emission dominating, note that
\begin{equation}
\tau_\alpha = z\theta^\alpha\,,
\end{equation}
or
\begin{equation}
\log\frac{1}{\tau_\alpha} = \log\frac{1}{z}+\alpha \log\frac{1}{\theta}\,,
\end{equation}
and so a fixed value of $\tau_\alpha$ corresponds to a straight line on the $(\log1/z,\log1/\theta)$ plane,  We can draw this as:

\begin{center}
\includegraphics[width=7.5cm]{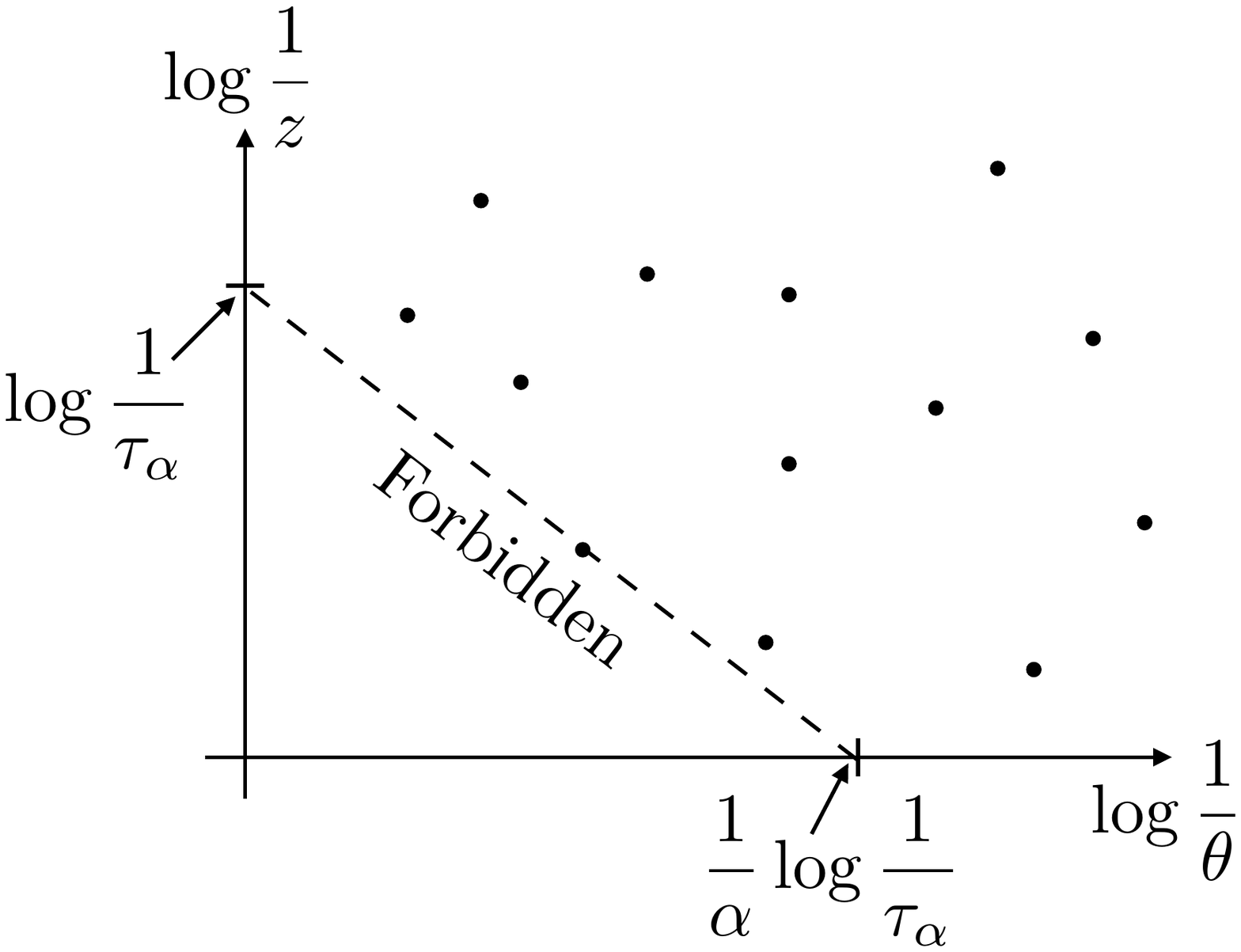}
\end{center}
We have drawn the abscissa and ordinate intercepts for the fixed value of $\tau_\alpha$ represented by the dotted line.  There is one emission on the dotted line, and all other emissions in the jet lie above the line.  Indeed, emissions below the dotted line are forbidden, as they would act to increase the value of $\tau_\alpha$ to be larger than what was measured.

So, for the measured value of $\tau_\alpha$, we must forbid all emissions at any point in the triangle below the dotted line.  To calculate the probability that the measured value of $\tau_\alpha$ is not larger than its value, we will do the following.  We must forbid emissions everywhere in the triangle, so let's isolate it and break it up into many pieces:

\begin{center}
\includegraphics[width=4cm]{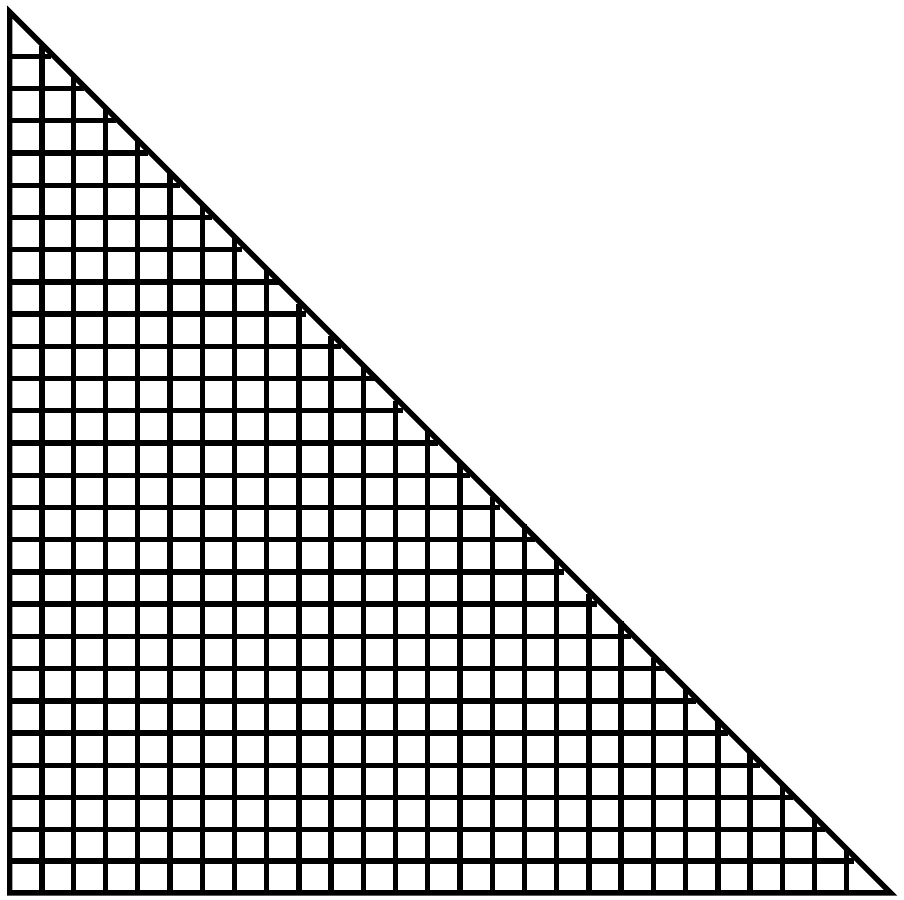}
\end{center}
There can be no emissions in any of the subregions.  This is an ``and'' statement in probability, so we must multiply the probability for no emissions in all regions together.  Let's break the triangle into $N$ equal area regions.  Then, the probability for an emission in one of the regions is uniform and equal to
\begin{equation}
\text{Probability for emission} = \frac{2\alpha_s}{\pi}C_F \frac{\includegraphics[width=0.5cm]{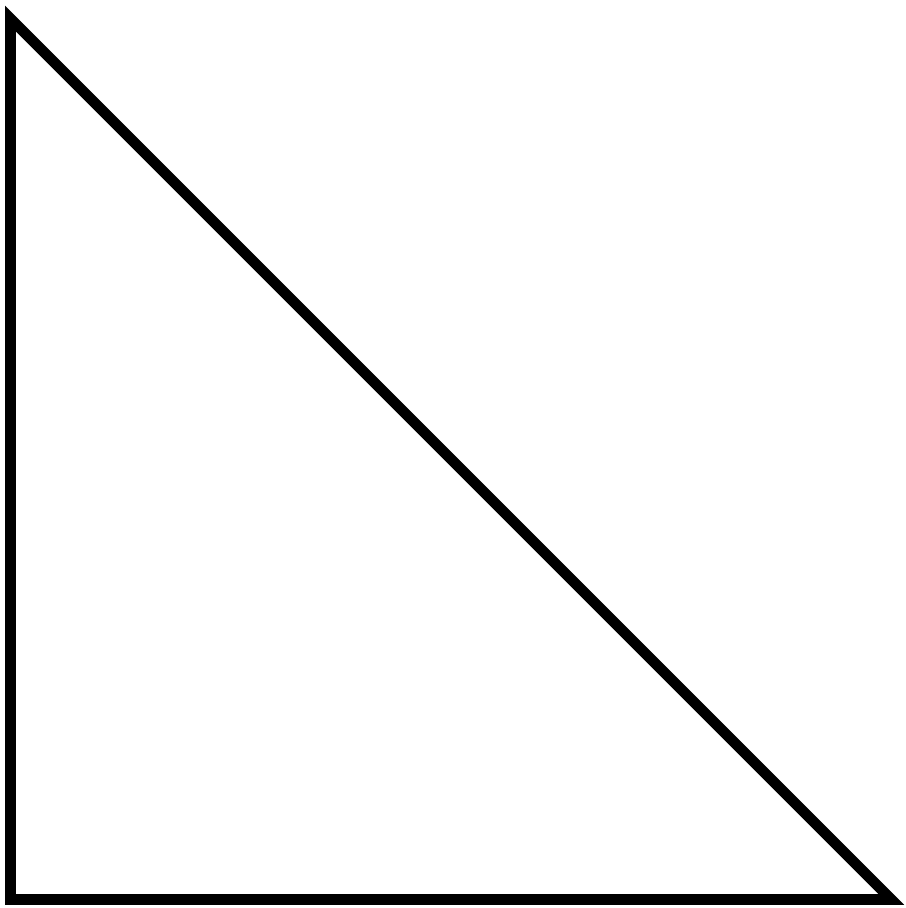}}{N}
\end{equation}
where \includegraphics[width=0.5cm]{triangle} is the area of the forbidden triangle:
\begin{equation}
\includegraphics[width=0.5cm]{triangle} = \frac{1}{2\alpha}\log^2\tau_\alpha\,.
\end{equation}
The, the probability for no emission in one small region is one minus this:
\begin{equation}
\text{Probability for no emission} = 1-\frac{2\alpha_s}{\pi}C_F \frac{\includegraphics[width=0.5cm]{triangle}}{N} = 1-\frac{\alpha_s}{\pi}\frac{C_F}{\alpha}\frac{\log^2\tau_\alpha}{N}\,.
\end{equation}
Then, the probability for no emission anywhere in the triangle is the product of probabilities of no emissions anywhere in each subregion:
\begin{equation}
P(\tau_\alpha \text{ less than measured value}) = \lim_{N\to\infty}\left(
1-\frac{\alpha_s}{\pi}\frac{C_F}{\alpha}\frac{\log^2\tau_\alpha}{N}
\right)^N = \exp\left[
-\frac{\alpha_s}{\pi}\frac{C_F}{\alpha}\log^2\tau_\alpha
\right]\,.
\end{equation}
The product transmogrifies into an exponential!  This exponential factor is called the Sudakov form factor \cite{Sudakov:1954sw}, and is simply a manifestation of the scale-invariant Poisson process of particle emission in high-energy QCD.

From a probability perspective, this Sudakov form factor is the cumulative probability distribution or CDF of $\tau_\alpha$.  To find the (differential) probability distribution function, we just differentiate:
\begin{equation}
p(\tau_\alpha) = \frac{d}{d\tau_\alpha}\exp\left[
-\frac{\alpha_s}{\pi}\frac{C_F}{\alpha}\log^2\tau_\alpha
\right] = -\frac{2\alpha_s}{\pi}\frac{C_F}{\alpha}\frac{\log\tau_\alpha}{\tau_\alpha}\exp\left[
-\frac{\alpha_s}{\pi}\frac{C_F}{\alpha}\log^2\tau_\alpha
\right]\,.
\end{equation}
This probability is normalized on $\tau_\alpha\in[0,1]$.  The issues with divergences with any fixed number of gluon emissions has been transformed into exponential suppression with the Sudakov form factor.

That's it for today---we'll use this intuition to understand aspects of machine learning next lecture.  Below are a couple of exercises.

\subsection*{Exercises}

\begin{enumerate}

\item Consider the measurement of two angularities, $\tau_\alpha$ and $\tau_\beta$, with, say $\alpha> \beta$.  Calculate the Sudakov form factor for two angularities, the joint probability distribution $p(\tau_\alpha,\tau_\beta)$.  Further, ensure that the joint probability distribution marginalizes to the correct single probability distributions.  That is,
\begin{equation}
\int_{\tau_0}^{\tau_1}d\tau_\beta\, p(\tau_\alpha,\tau_\beta) = p(\tau_\alpha)\,,
\end{equation}
for particular bounds $\tau_0< \tau_\beta < \tau_1$.  For a hint to this problem, see Ref.~\cite{Larkoski:2013paa}.

\item (This is an extension of Exercise 9.3 in Ref.~\cite{Larkoski:2019jnv}.)  The ALEPH experiment at the Large Electron-Positron Collider (LEP) measured the number of jets produced in $e^+e^-\to$ hadrons collision events.  The experiment counted $n$ jets, if, for every pair $i,j$ of jets the following inequality is satisfied:
\begin{equation}
2\min[E_i^2,E_j^2](1-\cos\theta_{ij}) > y_{\text{cut}} E_{\text{cm}}^2\,,
\end{equation}
for $y_\text{cut}<1$, $E_i$ is the energy of jet $i$, $\theta_{ij}$ is the angle between jets $i$ and $j$ and $E_\text{cm}$ is the center-of-mass collision energy.  In the soft and collinear limits, determine the probability $p_n$ for observing $n$ jets, as a function of $y_\text{cut}$.

Note that the minimum number of jets is 2 ($e^+e^- \to q\bar q$) and gluons can be emitted from either the quark or the anti-quark.  Compare your result to figure 7 of Ref.~\cite{Heister:2003aj}.  What value of $\alpha_s$ fits the data the best?  This fit is imperfect because we're omitting a lot of important physics, but it will be qualitatively close.

\end{enumerate}

\section{Lecture 2: Machine Learning for Quark vs.~Gluon Discrimination}

Welcome back to the second lecture on jets and machine learning.  In this lecture, we're going to think like a machine to understand the problem of binary discrimination, or, techniques for distinguishing two samples mixed in an ensemble.  Before we get to that, I want to wrap up a couple of points from the previous lecture that we will use to great effect this lecture.  The first is the property of infrared and collinear safety of the angularities.  We had found that the Sudakov form factor for the angularity measured on a quark jet took the form:
\begin{equation}
\Sigma_q(\tau_\alpha) = \exp\left[
-\frac{\alpha_s}{\pi}\frac{C_F}{\alpha}\log^2\tau_\alpha
\right]\,,
\end{equation}
for $\tau\in[0,1]$ and $\alpha > 0$.  Recall that $\alpha$ was the angular exponent of the angularity:
\begin{equation}
\tau_\alpha = \sum_{i\in J}z_i \theta_i^\alpha\,,
\end{equation}
and I said that it must be positive for infrared and collinear safety.  Using our axiom of scale invariance of QCD, we were lead to a probability for single gluon emission that diverged in the soft (= low energy) and collinear limits.  When all degenerate configurations of emitted gluons are summed up, we generate the Sudakov form factor, which is finite by the KLN theorem.  However, the Sudakov form factor is a series with terms to all orders in the coupling $\alpha_s$.  What if we just want a description of QCD jets to a fixed order in $\alpha_s$?  That is, a description as provided by some collection of Feynman diagrams?  Well, for the result to be sensible (i.e., finite) even though the fundamental probability distribution diverges in the soft and/or collinear limits, requires a delicate property of the observable that we choose to measure on the jet.  In particular, for the distribution of an observable to be finite almost everywhere in its domain requires that the soft and collinear limits of that observable map to a unique value.  That is, there is a single value of the observable for which all divergences from soft or collinear gluon emission are located.  Another way to state this criteria is that exactly 0 energy or exactly collinear gluons do not affect the value of the observable \cite{Ellis:1991qj}.  Such an observable for which this is true is called ``infrared and collinear safe'' or IRC safe.  Isolating all divergences to a single value of the observable means that away from that value, everything is well-defined and finite.

The angularities are IRC safe because soft and collinear gluons do not contribute to $\tau_\alpha$, for $\alpha > 0$.  However, not all possible observable or questions you can ask of a jet are IRC safe.  Perhaps the canonical example of non-IRC safety is that of multiplicity, or the number of particles in a jet.  A jet could consist of a single, bare quark so multiplicity would be 1.  However, say that a quark emits an exactly collinear gluon; now multiplicity would be 2.  However, this exactly collinear emission is degenerate to the bare quark, and violates the assumptions of KLN for finiteness.  So, multiplicity is not IRC safe.

Another thing to address now is what the Sudakov form factor is for a high-energy gluon jet.  From what we have discussed thus far, the only difference between a quark and gluon is the color that either carry.  (Implicitly also their spin is importantly different, but we won't pursue that more here.)  A quark is in the fundamental representation of SU(3) color, and so the number of colors it can share with a soft or collinear gluon is controlled by the fundamental quadratic Casimir, $C_F = 4/3$.  Gluons, by contrast, live in the adjoint representation of SU(3) color, and can share more color with soft or collinear gluons than can quarks.  The adjoint quadratic Casimir $C_A = 3$ in QCD, so this has consequences for the gluon Sudakov factor and a heuristic for understanding properties of QCD jets.  The gluon Sudakov is found by replacing $C_F \to C_A$ in the quark Sudakov:
\begin{equation}
\Sigma_g(\tau_\alpha) = \exp\left[
-\frac{\alpha_s}{\pi}\frac{C_A}{\alpha}\log^2\tau_\alpha
\right]\,.
\end{equation}
Because $C_A > C_F$, gluons are more likely to emit soft/collinear gluons, so it is more difficult for a gluon jet to have a very small value of $\tau_\alpha$.  Hence, the Sudakov provides a greater exponential suppression for gluons than quarks.

With those points established, I now want to pivot to discussing machine learning in a very restricted (and biased!) manner.  Machine learning is exploding as a discipline in particle physics \cite{Larkoski:2017jix,Guest:2018yhq,Radovic:2018dip,Albertsson:2018maf,Carleo:2019ptp}, so there is no hope for me to discuss broadly how it is employed.  However, my biased viewpoint is from that of a theorist who is selfishly interested in learning more about Nature.  So how can we use machine learning to learn more aa flesh-and-blood humans?  Let's first define what we are working with.

My theorist definition of a machine (or neural network or any fancy computer science algorithm) is the following.  A machine is a black box which takes in input and returns output:

\begin{center}
\includegraphics[width=4.5cm]{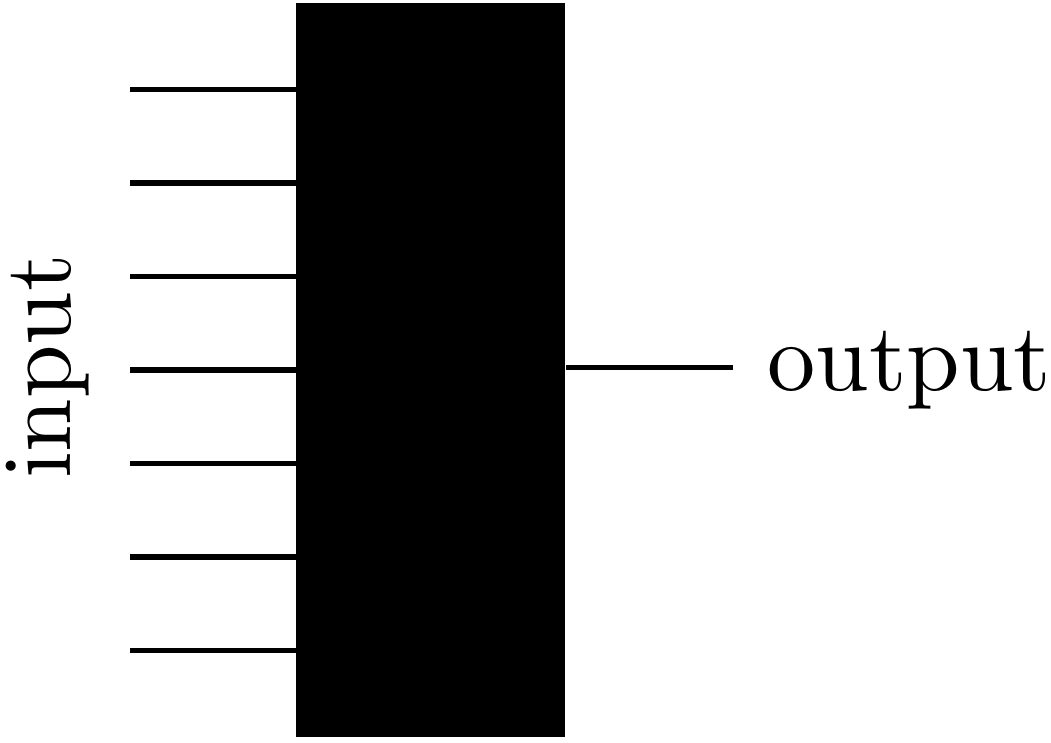}
\end{center}
By ``black box'', I mean that the machine performs some manipulation on the input to produce the output, but the way in which it does it is unknowable to me.  Don't worry; it will turn out fine that we don't know what the machine does.  I have also illustrated the input as multiple entries and the output as a single result.  That is, we consider the input to be some $n$ dimensional vector $\vec x$ and the output to be a single number $g$.  So all the machine is is a function of the input:
\begin{equation}
\text{Machine: }\mathbb{R}^n \to \mathbb{R}\,,
\end{equation}
which can be represented by the function $g\equiv g(\vec x)$.  One can consider more general inputs and outputs, but we will keep it restricted to this scalar function case, again for simplicity.

The first result that allows us to learn anything at all is the universal approximation theorem \cite{cybenko,hornik,leshno}.  For our purposes, the statement of the universal approximation theorem is the following.  A sufficiently large and powerful machine can output an arbitrary function of the inputs.  Concretely, let's say that the input data are entries of the vector $\vec x$, and the output of the machine is the function $g(\vec x)$:

\begin{center}
\includegraphics[width=4cm]{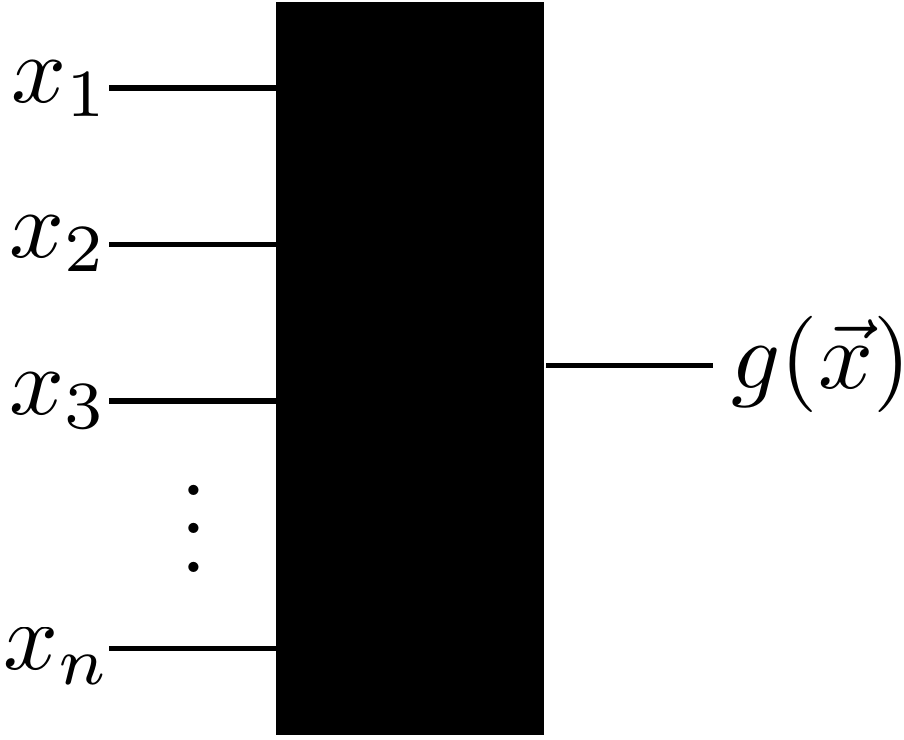}
\end{center}
The universal approximation theorem states that we can modify the inputs to (almost) any collection of functions $\{f_i(\vec x)\}_{i=1}^n$ and the machine will return (or ``learn'') the same function $g(\vec x)$:

\begin{center}
\includegraphics[width=4.5cm]{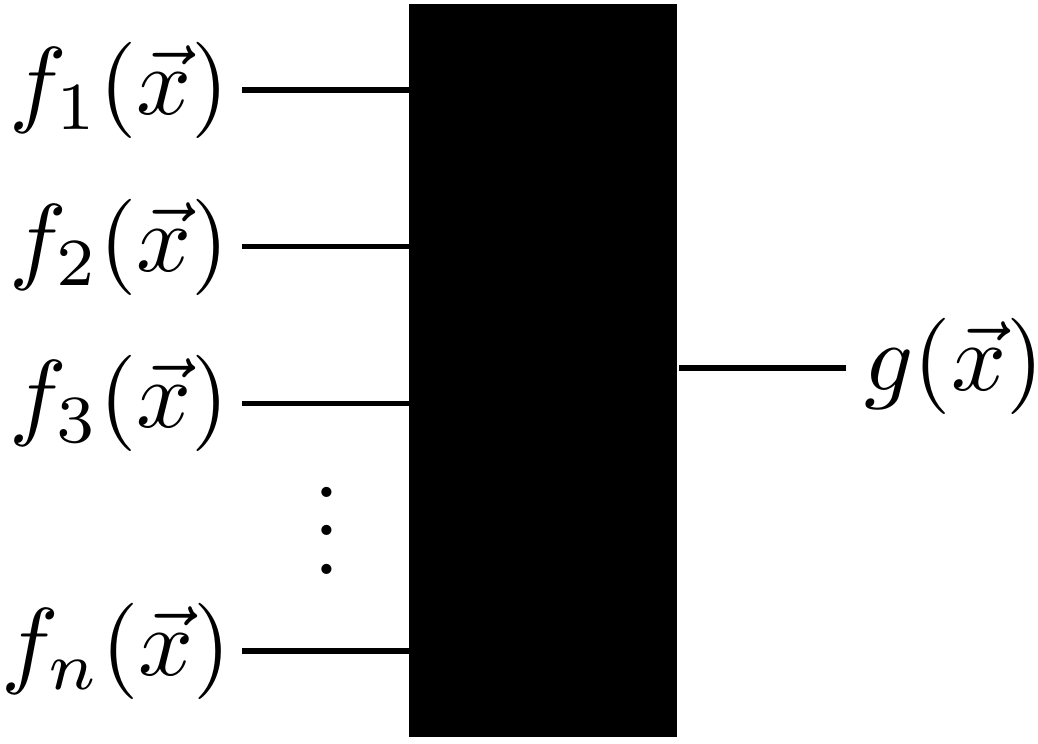}
\end{center}
As long as the collection of $f_i$ functions contains the same total information as the $\vec x$ input data, the machine will always learn $g(\vec x)$.  So, the universal approximation theorem states that we can choose a convenient form of input data, and so we can optimize for a form we understand well.

Now, a lot of machine learning practitioners will say that the universal approximation theorem is less than unuseful because the actual behavior of the machine depends so sensitively on precise implementation and architecture.  However, I again want to emphasize that we are not interested here in determining how a particular machine performs; we want to think like a machine to learn something about Nature.

The specific task we would like the machine to perform is binary discrimination, or, signal versus background discrimination.  The formulation of a binary discrimination problem is that we want the machine to separate, as efficiently as possible, signal from background events in a mixed ensemble.  What we mean by ``signal'' and ``background'' is that signal events, or a collection of the input data, are drawn from the probability distribution $p_s(\vec x)$, while background events are drawn from the probability distribution $p_b(\vec x)$.  Given a collection of identified signal and background events, the machine learns what the probability distributions are, and correspondingly outputs the probability with which it believes that an individual event is signal-type or background-type.

It doesn't do this separation blindly; we know the optimal binary discriminant from the result of the Neyman-Pearson lemma \cite{Neyman:1933wgr}.  In the 1930s, Jerzy Neyman and Egon Pearson proved that the likelihood ratio is the optimal binary discriminant.  The likelihood ratio ${\cal L}$ is simply the ratio of signal to background probability distributions:
\begin{equation}
{\cal L} = \frac{p_s(\vec x)}{p_b(\vec x)}\,.
\end{equation}
The likelihood naturally and beautifully separates signal and background, to the maximal extent provided by the probability distributions.  If ${\cal L}\to 0$, then the background probability is large compared to signal, and vice-versa for ${\cal L}\to \infty$.  Thus,
\begin{align}
&{\cal L}\to 0\qquad \Rightarrow \qquad \text{pure background}\,,\\
&{\cal L}\to \infty\qquad \Rightarrow \qquad \text{pure signal}\,.
\end{align}
Additionally, a larger class of quantities than strictly just the likelihood are equal in discrimination power.  Any function monotonic in the likelihood ${\cal L}$ is equivalent in discrimination power.  This monotonic function can be exploited to simplify the range of values that the discrimination observable assumes.  For example, the function
\begin{equation}
h({\cal L}) = \frac{{\cal L}}{1+{\cal L}} = \frac{p_s(\vec x)}{p_s(\vec x)+p_b(\vec x)}\,,
\end{equation}
is monotonic in ${\cal L}\in[0,\infty)$, but nicely maps to the domain $h({\cal L})\in[0,1]$.  In particular:
\begin{align}
&h({\cal L}\to 0)\to 0\qquad \Rightarrow \qquad \text{pure background}\,,\\
&h({\cal L}\to \infty)\to 1\qquad \Rightarrow \qquad \text{pure signal}\,.
\end{align}
Again, there's nothing special about this $h({\cal L})$, but it can be a nice object to consider.

So, for the case at hand, our machine takes in a collection of input data $\vec x$ for signal and background events drawn from probability distributions $p_s(\vec x)$ and $p_b(\vec x)$, respectively, and outputs the likelihood ratio, or a monotonic function of it:

\begin{center}
\includegraphics[width=5cm]{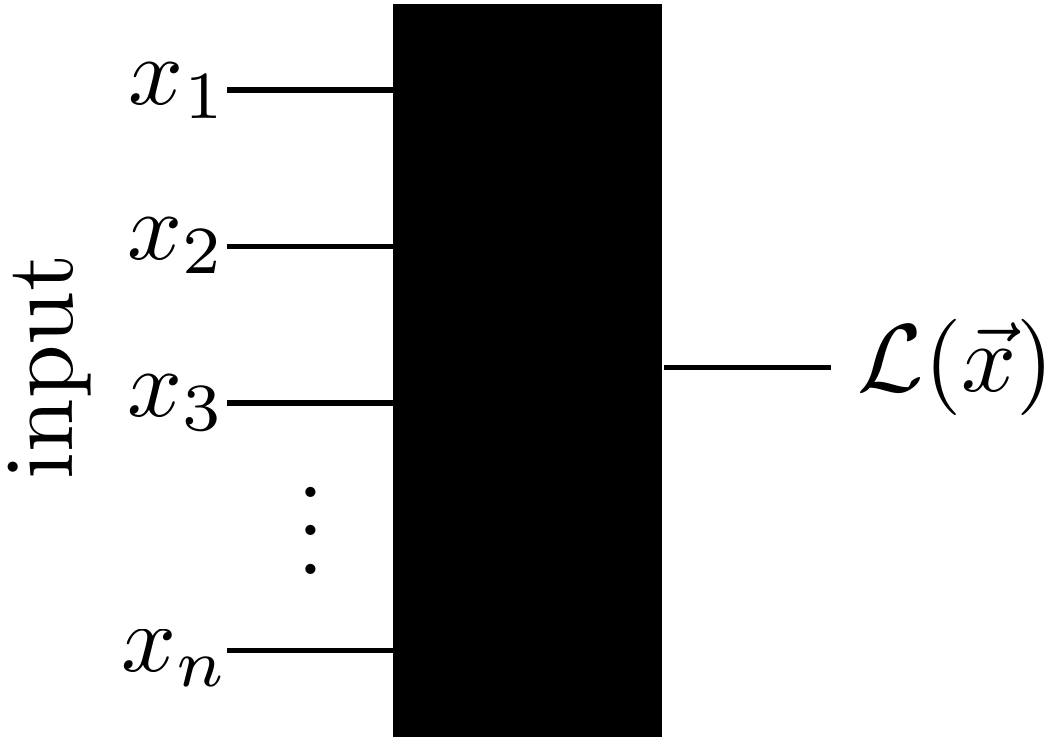}
\end{center}
With that set-up, let's now describe the physical system we would like to learn about.  Our goal in the rest of this lecture is to learn something about the likelihood ratio for the problem of discrimination of jets initiated by high-energy quarks versus those jets initiated by high-energy gluons.  The input data that we will use in our machine is the collection of four-momenta of particles in quark or gluon jets.  Let's call this set of momenta $\{p_i\}_{i\in J}$, and the quark and gluon probability distributions are thus:
\begin{equation}
p_q\left(\{p_i\}_{i\in J}\right)\,, \qquad
p_g\left(\{p_i\}_{i\in J}\right)\,,
\end{equation}
where $J$ is the jet of interest (i.e., the collection of soft and collinear particles).  The likelihood ratio is then
\begin{equation}
{\cal L} = \frac{p_g\left(\{p_i\}_{i\in J}\right)}{p_q\left(\{p_i\}_{i\in J}\right)}\,.
\end{equation}
So, there we go.  We're done, right?  While true, there is almost zero information in these statements; specifically, we know essentially nothing about the functional form of ${\cal L}$ for quark and gluon discrimination.  Can we learn any information that will help us in our task?

The first problem with this formulation is the explicit use of particle four-vectors as the input data.  This is a rather poor way to organize the information in a jet because, by the soft and collinear singularities of QCD, a substantial amount of particles will have (nearly) degenerate momenta, which is challenging to interpret theoretically so we humans can actually learn something.  So, using the universal approximation theorem, let's see if we can reorganize the information contained in four-vectors into a much more useful, and human-interpretable, form.

First, let's see what we are dealing with and what the dimension of the input space actually is.  A generic four-vector has four real components, so the four-vectors of $N$ particles (na\"ively) spans some real $4N$-dimensional space.  However, as real particles, their momenta are all on-shell.  For simplicity (but no other reason) let's assume that all particles are massless.  As such, an on-shell, massless four-vector actually only has 3 degrees of freedom, so the space of input momenta is only $3N$ real dimensions.  Additionally, total energy and momentum are conserved, which imposes 4 further linear constraints on all momenta.  Thus, there are $3N-4$ degrees of freedom to completely define the four-momenta of $N$ particles in a jet (assuming on-shellness and total momentum conservation).  So, we just need to identify $3N-4$ functions of the particles' momenta appropriately and we can use the universal approximation theorem to claim that our machine would find the same likelihood ratio.

So, what functions of momenta should we use?  This is a matter of taste, but if we want to exploit our perturbative understanding of QCD to the problem at hand, then we would want these functions to be infrared and collinear safe observables.  What $3N-4$ IRC safe observables should we use?  This is much more of a matter of taste, and there are many possible answers, but here we will just consider the $N$-subjettiness observables \cite{Thaler:2010tr,Thaler:2011gf}.  $N$-subjettiness is a class of IRC safe observables that extends the angularities to resolve $N$ prongs in a jet.  As a function of energy fractions $z_i$ and angles, $N$-subjettiness $\tau_N^{(\alpha)}$ is defined to be:
\begin{equation}
\tau_N^{(\alpha)} = \sum_{i\in J}z_i\, \min\left\{
\theta_{i1}^\alpha\,,\theta_{i2}^\alpha\,,\dotsc\,, \theta_{iN}^\alpha
\right\}\,,
\end{equation}
and $\alpha > 0$.  Here, $\theta_{iK}$ is the angle between particle $i$'s momentum and axis $K$ in the jet.  Defining $N$-subjettiness requires placing $N$ axes in the jet, nominally in the directions of dominant energy flow.  For example, consider $2$-subjettiness measured on a jet with two hard particles, 1 and 2, and one soft particle 3.  The two axes would, for example, align with particles 1 and 2 and only particle 3 would contribute to $\tau_2^{(\alpha)}$:
\begin{align}
\raisebox{-1.5cm}{\includegraphics[width=3cm]{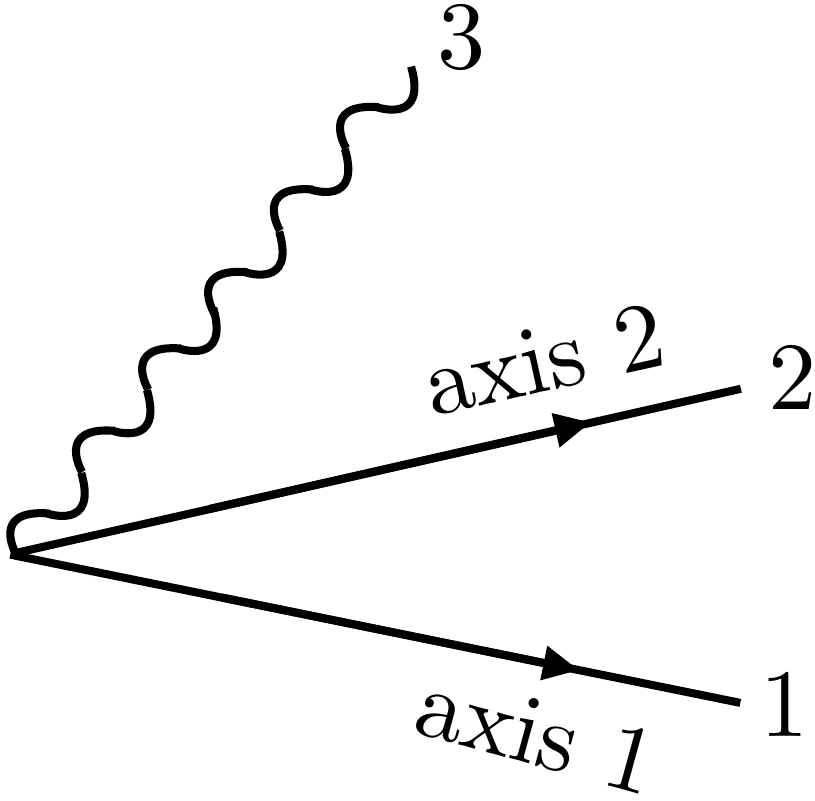}}\qquad  \Rightarrow \qquad \tau_2^{(\alpha)} &=\sum_{i\in J}z_i\, \min\{\theta_{i1}^\alpha,\theta_{i2}^\alpha\}\\
  &=z_3 \min\{\theta_{31}^\alpha,\theta_{32}^\alpha\}\,.\nonumber
\end{align}
Angularities are 1-subjettiness, and IRC safety of $N$-subjettinesses essentially follows from IRC safety of angularities (with some caveats regarding axes choice).  $N$-subjettinesses have the added benefit that they are additive, and so multiple soft and collinear emissions in the jet generates a Sudakov form factor, exactly as we observed for angularities.

So, we already know a lot about the space of $N$-subjettinesses, so we can choose $3N-4$ of them to resolve the four-momenta of $N$ particles in the jet.  Because time is short, I won't go into explicit details about what that collection of $N$-subjettinesses should be to ensure that they have the same information as the collection of four-vectors.  See Ref.~\cite{Datta:2017rhs} for all the details.

Now we're cooking.  What properties of the likelihood ratio ${\cal L}$ for quark versus gluon discrimination can we learn using the $N$-subjettiness variables as inputs to our machine (which in this case is our brains!)?  I'll present a simplified argument here, and more details can be found in Ref.~\cite{Larkoski:2019nwj}.  Let's just imagine, for simplicity, our entire input space was just that of $\tau_1^{(2)}$ and $\tau_2^{(2)}$, the 1- and 2-subjettinesses with angular exponents equal to 2:
\begin{align}
&\tau_1^{(2)} =\sum_{i\in J}z_i \theta_i^2\,, &\tau_2^{(2)} = \sum_{i\in J}z_i\, \min\{\theta_{i1}^2,\theta_{i2}^2\}\,.
\end{align}
Note that both $\tau_1^{(2)},\tau_2^{(2)}>0$ and $\tau_1^{(2)}>\tau_2^{(2)}$ because with two axes in the jet the distance to any particle to those axes is less than or equal to the distance to a single axis in the center of the jet:

\begin{center}
\includegraphics[width=8.5cm]{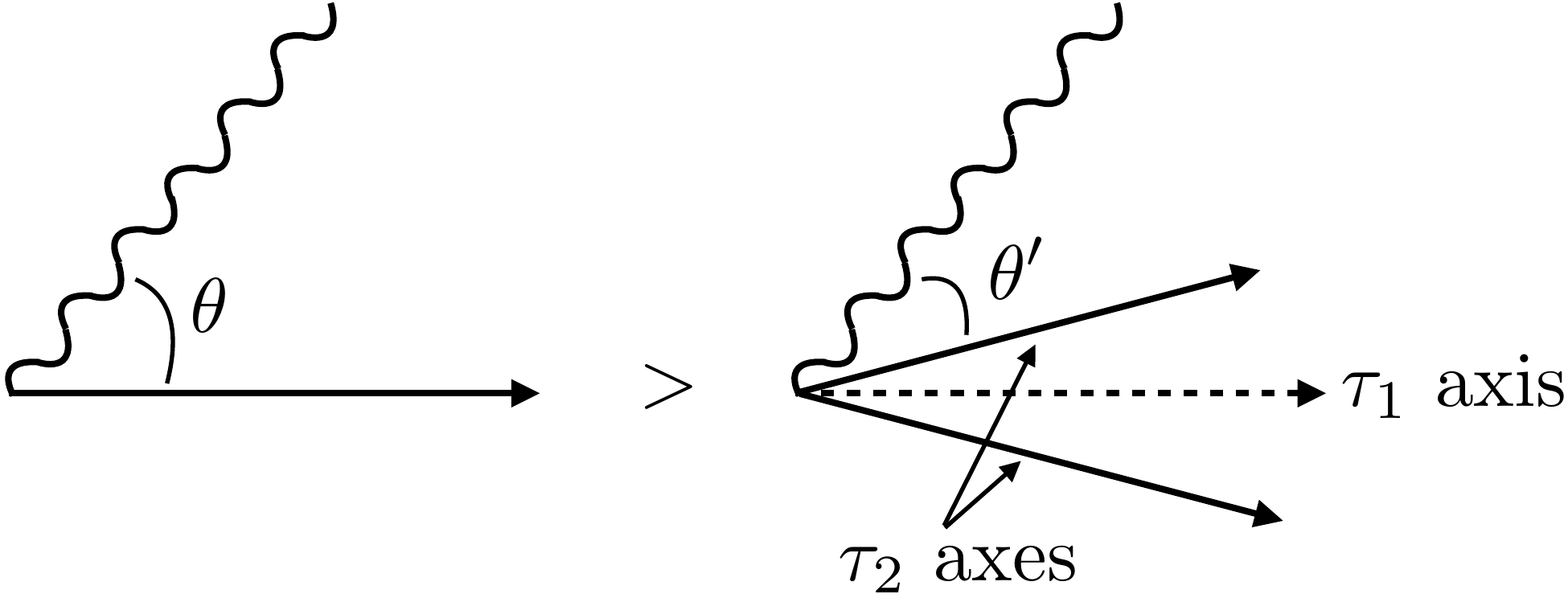}
\end{center}
Thus our phase space defined by measuring $\tau_1^{(2)}$ and $\tau_2^{(2)}$ looks like:

\begin{center}
\includegraphics[width=4cm]{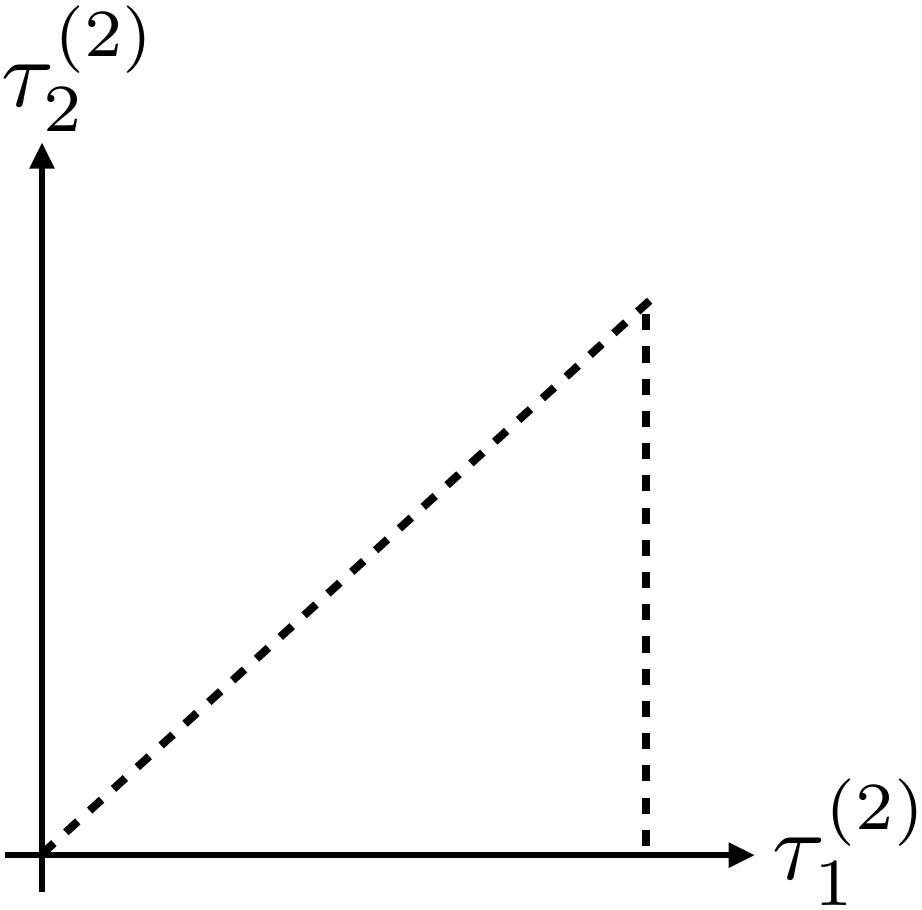}
\end{center}
Phase space is the triangular region bounded by the $\tau_1^{(2)}$ axis and the dashed lines.

To go further, we need two things: (1) identification of the soft/collinear region and (2) the form of the likelihood on this space.  (1) is easy enough to answer: because $\tau_N^{(\alpha)}>0$ and is IRC safe, the $\tau_N^{(\alpha)}\to 0$ limit is the soft/collinear divergent limit.  On our phase space, this is just the region near the $\tau_1^{(\alpha)}$ axis:

\begin{center}
\includegraphics[width=4cm]{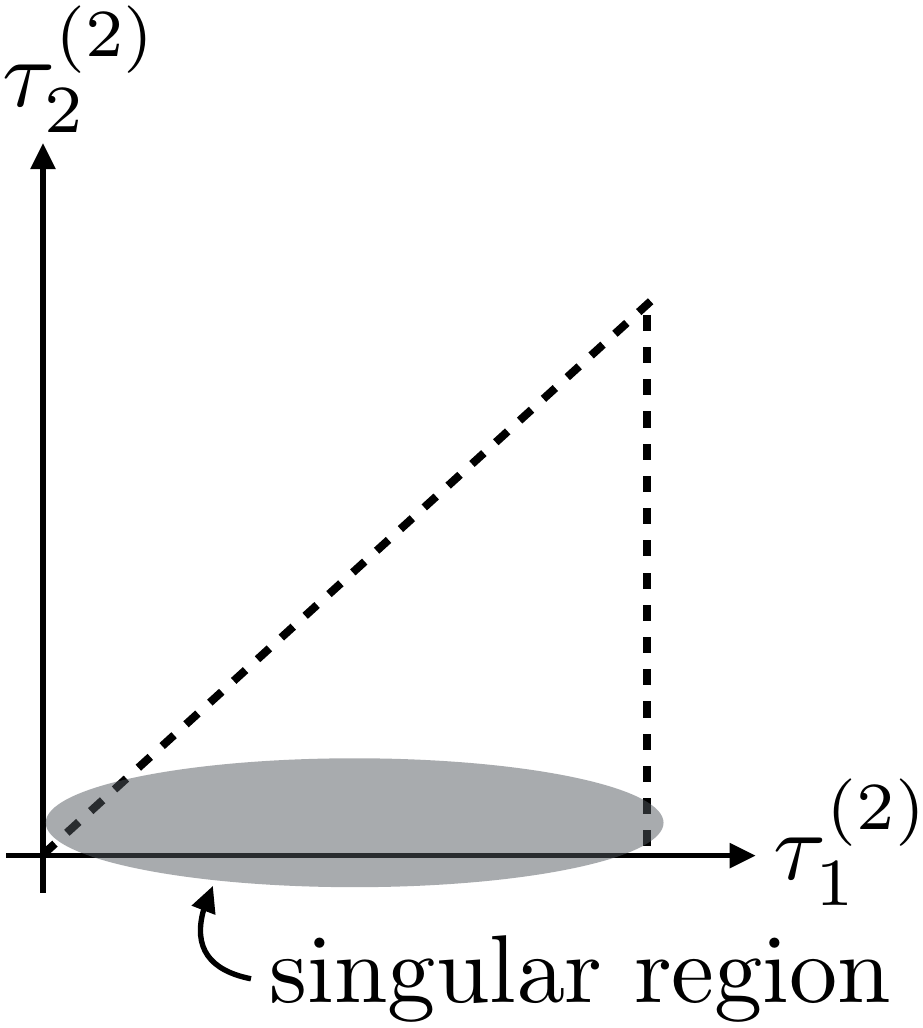}
\end{center}

Okay, let's see why this observation is so important.  The likelihood ratio of quark and gluon probability distributions on this space is
\begin{equation}
{\cal L}(\tau_1^{(\alpha)},\tau_2^{(\alpha)}) = \frac{p_g(\tau_1^{(\alpha)},\tau_2^{(\alpha)})}{p_q(\tau_1^{(\alpha)},\tau_2^{(\alpha)})}\,.
\end{equation}
Do we have any information as to what the functional form of this ratio might be?  Yes, from the Sudakovs!  By the IRC safety (and additivity) of $N$-subjettiness, the quark and gluon probabilities have Sudakov factors of the form
\begin{align}
&p_q(\tau_1^{(\alpha)},\tau_2^{(\alpha)}) \sim e^{-\alpha_s C_F f(\tau_1^{(\alpha)},\tau_2^{(\alpha)})}\,,
&p_g(\tau_1^{(\alpha)},\tau_2^{(\alpha)}) \sim e^{-\alpha_s C_A f(\tau_1^{(\alpha)},\tau_2^{(\alpha)})}\,.
\end{align}
Now, without an explicit calculation, we don't know what the functional form of $f(\tau_1^{(\alpha)},\tau_2^{(\alpha)})$ is, but, by IRC safety of $N$-subjettiness, we know its behavior in limits.  The soft/collinear limit corresponds to $\tau_2^{(2)}\to 0$ (with $\tau_1^{(2)}>\tau_2^{(2)}$), and in this limit, the Sudakov factor exponentially suppresses the probability distributions.  With this exponential form, we then must have that
\begin{equation}
f(\tau_1^{(\alpha)},\tau_2^{(\alpha)}\to0)\to\infty\,.
\end{equation}

Now, the form of the likelihood ratio for quark versus gluon jet discrimination is
\begin{equation}
{\cal L}(\tau_1^{(\alpha)},\tau_2^{(\alpha)}) = \frac{p_g(\tau_1^{(\alpha)},\tau_2^{(\alpha)})}{p_q(\tau_1^{(\alpha)},\tau_2^{(\alpha)})} \sim e^{-\alpha_s (C_A-C_F) f(\tau_1^{(\alpha)},\tau_2^{(\alpha)})}\,.
\end{equation}
Because $C_A>C_F$ in QCD, we still have that the likelihood vanishes in the singular, $\tau_2^{(2)}\to 0$ limit:
\begin{equation}
{\cal L}(\tau_1^{(\alpha)},\tau_2^{(\alpha)}\to 0)\to 0\,.
\end{equation}
However, the entire region where $\tau_2^{(2)}\to 0$ is the soft/collinear limit in which a fixed-order description of jets diverges.  In this entire singular region, the likelihood takes a unique value: ${\cal L} = 0$.  Thus, the likelihood for quark versus gluon discrimination is itself IRC safe, from our earlier discussion.  Out of all possible functions of input particle momenta, the likelihood for this problem is IRC safe, which is a strong constraint on its form.  Thus, we learn that if you want to distinguish quark-flavor from gluon-flavor jets, a good place to start is to use an IRC safe observable.

Again, I want to emphasize that we, humans, learned something about QCD by thinking like a machine.  What else might we learn with this approach?  I hope you can find something new!

\subsection*{Exercises}

\begin{enumerate}

\item Just consider the angularities for discrimination of quark vs.~gluon jets.

\begin{enumerate}

\item What is the likelihood ratio for the probability distributions $p_g(\tau_\alpha)$ and $p_q(\tau_\alpha)$?

\item What is the distribution of this likelihood ${\cal L}$ on quark jets and gluon jets, $p_q({\cal L})$ and $p_g({\cal L})$?

\item The receiver operating characteristic curve (ROC) quantifies the ``strength'' of separation power of the likelihood, as a function of a cut on the likelihood.  If the cumulative distributions of the likelihood for quarks and gluons are
\begin{align}
&\Sigma_q({\cal L}) = \int_0^{\cal L}d{\cal L}'\, p_q({\cal L}')\,,
&\Sigma_g({\cal L}) = \int_0^{\cal L}d{\cal L}'\, p_g({\cal L}')\,,
\end{align}
the ROC curve is defined to be
\begin{equation}
\text{ROC}(x) = \Sigma_g(\Sigma_q^{-1}(x))\,,
\end{equation}
where $\Sigma_q^{-1}(x)$ is the inverse of the quark's cumulative distribution function.  What is the ROC, as a function of the quark fraction $x$?

\item The area under the ROC curve (AUC) is also an interesting discrimination metric, often used by a (real) machine in a gradient descent algorithm.  What is the AUC for the ROC curve in part (c)?

\end{enumerate}

\item Consider the number of jets as defined by the procedure introduced in Exercise (2) of Lecture 1.  Consider that procedure measured on $e^+e^-\to q\bar q+X$ and $e^+e^-\to gg+X$ events, where $X$ is any other hadronic activity.  Using the discrete probability distribution $p_n$ for the quark and gluon final states, determine the likelihood ratio, ROC curve and AUC for this number of jets observable for discrimination of the quark from gluon final states, as a function of the parameter $y_\text{cut}$.  Does the AUC for this observable ever correspond to better discrimination to that for $\tau_\alpha$ from Exercise (1) above?

\end{enumerate}

\section*{Acknowledgments} 

I thank Artur Apresyan and Stefan Hoeche for inviting me to lecture at the 2020 HCPSS.  The school was a success due to their hard work, even with the extremely challenging times.  I also thank Artur, Stefan, and others who keep inviting me to talk about machine learning because of my skepticism, wariness and outright refusal to work on programming of a neural network.

\end{document}